\documentclass{elsart}
\usepackage{graphicx}
\usepackage{amsfonts}
\usepackage{amsmath}
\usepackage{amssymb}

\def\Journal#1#2#3#4#5{{#1}, {#2} {#3} (#5) #4.}

\newcommand{\KI}{K_0}
\newcommand{\HI}{H_0}
\newcommand{\SI}{S_0}

\newcommand{\EE}{\mathbb{E}}

\newcommand{\uu}{\mathbf 1}
\newcommand{\FF}{\mathcal{F}\/}

\newcommand{\tpm}{\tau^{\pm}}

\newcommand{\BS}{BS }
\newcommand{\betap}{\alpha}
\newcommand{\betam}{\beta}

\begin{document}
\begin{frontmatter}

\title{Perpetual American options within CTRW's}
\author{Miquel Montero}
\address{Departament de F\'{\i}sica Fonamental, Universitat de
Barcelona, Diagonal 647, E-08028 Barcelona, Spain}
\ead{miquel.montero@ub.edu}

\date{\today}

\begin{abstract}
Continuous-time random walks are a well suited tool for the description of market behaviour at the smallest scale: the tick-to-tick evolution. We will apply this kind of market model to the valuation of {\it perpetual\/} American options: derivatives with no maturity that can be exercised at any time. Our approach leads to option prices that fulfil financial formulas when canonical assumptions on the dynamics governing the process are made, but it is still suitable for more exotic market conditions.   
\end{abstract}


\end{frontmatter}

\section{Introduction}
\label{Sect_Intro}

Since their introduction in 1965 by Montroll and Weiss~\cite{MW65}, continuous-time random walks (CTRW's) have been applied to virtually any field in which one wants to capture the smallest-scale properties of a given random system, and finance is not an exception. In fact, an intense activity in this area have been developed in recent years, as it can be inferred from the extensive (and exhaustive) list of references included in the review that Scalas~\cite{ES06} published in 2006. However, most of the work involving CTRW models was intended to reproduce empirical statistical properties, like probability density function (PDF) of returns~\cite{MMW03}, or the mean exit time of the process out of a given region~\cite{MMP05,MPMLMM05}. The amount of such articles that deal with problems with some application on option pricing, as in Ref.~\cite{MM07}, is not so large instead. 

In this article we present pricing expressions for several perpetual American options. The results were obtained by applying standard CTRW techniques on a market model whose time evolution is step-like. The fact is that {\it option pricing when underlying stock returns are discontinuous\/} is not a new problem at all~\cite{RCM76}, and it is under active research nowadays in mathematical finance~\cite{CT04}. The main differences between these approaches are twofold: we do not assume either that the process evolution is diffusive between changes, CTRW collects {\bf all} the stochastic (tick-by-tick) dynamics, nor transactions constitute necessarily a {\bf Poisson process}. 
This might allow us to explore inefficient market models. When a risk-neutral market evolution can be defined we obtain fair option prices, from which classical Black-Scholes (BS) results can be recovered.

\section{The process}
\label{Sect_process}

\begin{figure}[hbtp]
  \begin{center}
\includegraphics[width=0.80\textwidth,keepaspectratio=true]{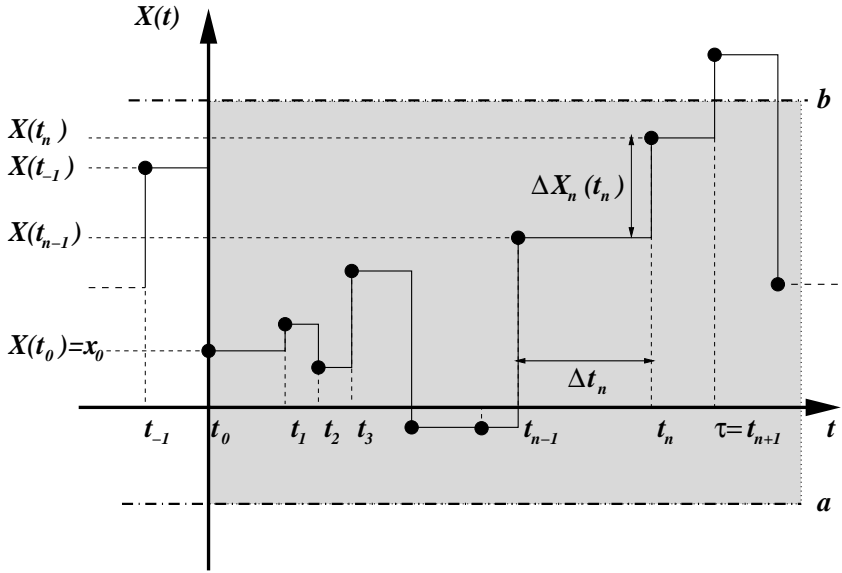} \caption{A sample trajectory of the $X(t)$ process along with the corresponding value of the random variable $\tau$, the exercise time.} \label{model}
\end{center}
\end{figure}

We will model the market evolution of stock prices, $S(t)$, through the logarithmic return $X(t)=\ln[S(t)]$. We will assume that the stochastic process follows the most common version of the CTRW: $X(t)$ shows a series of random increments or jumps at random times $\cdots,t_{-1},t_0,t_1,t_2,\cdots,t_n,\cdots$, the transaction times, remaining constant between these jumps. Therefore, the resulting trajectory consists of a series of steps as shown in Fig.~\ref{model}. We will assume that waiting times, $\Delta t_n=t_n-t_{n-1}$, and random returns, $\Delta X_n(t_n)=X(t_n)-X(t_{n-1})$, are (mutually) independent and identically distributed random variables, described by two PDF's which we will denote by $\psi(t)$ and $h(x)$ respectively. 

A key magnitude in the CTRW formalism is the propagator, $p(x-x_0,t-t_0)$, 
the PDF of $X(t)=x$ when one knows the value of the process right after a jump, $X(t_0)=x_0$.
Under our previous assumptions the propagator fulfils a renewal equation~\cite{ES06} 
that can be readily solved in the Fourier-Laplace domain: 
\begin{equation}
  \hat{\tilde p}(\omega ,s) = \frac{1}{s}\frac{{1 - \hat\psi (s)}}{{1 - \tilde h(\omega )\hat\psi (s)}},
\label{hat_tilde}
\end{equation} 
where tildes and hats stand for Fourier and Laplace transforms respectively.

\section{Option pricing}
\label{Sect_pricing}

Options are contracts between two parties, sold by one party to another, that give the buyer the right, but not the obligation, to buy (call) or sell (put) shares of the underlying stock at some prearranged price, the {\it strike\/} price $K$, within a certain period or on a specific date, the maturity or expiration time $T$. Therefore options are contingent claims with their present value determined by the discounted value of the expected profit under a {\bf risk-neutral measure}: 
\begin{equation}
C(\SI,t_0)=\EE [P(S,\tau) e^{-r(\tau-t_0)} \uu_{\tau \leqslant  T}|\FF(t_0)],
\label{option_price}
\end{equation}
where $r$ is the risk-free interest rate, $\tau$ is the actual exercise time, $P(S,t)$ is the pay-off function, and $\FF(t)$ condenses the available information up to time $t$.
We will concentrate our attention on 
pay-off functions of the current value of the asset, like in the case of {\it vanilla\/} calls (+) and puts (-) where
\begin{equation}
P^{\pm}(S,\tau)=
\pm[S(\tau)-K] \uu_{S(\tau) \gtrless  K},
\label{payv}
\end{equation}
and $K$ is constant. Another broadly used pay-off is 
$P^{\pm}(S,\tau)=\uu_{S(\tau) \gtrless  \KI}$,
which corresponds to binary or digital call and put options. 

When the option can be exercised at the end of the contract lifetime solely, $\tau=T$, the option is said to be European. If the option can be exercised at any time before expiration it is called American, and $\tau$ becomes a stochastic magnitude. Since the contract becomes worthless after maturity the option holder must find, under this constraint, the optimal exercise boundary, $H(t)$: the stock price above (below) which it is better to exercise the call (put) than to keep the option alive. 
We will define the exercise time $\tpm$ as the first time the underlying crosses the threshold given that at present time, $t_0$, the spot price of the asset, $\SI$, lies in the proper side of the boundary:  
\begin{equation*}
\tpm=\min\left\{t > t_0; S(t) \gtrless H^{\pm}(t)|\SI \lesseqgtr \HI^{\pm}\right\}.
\end{equation*}
Note that here the boundary function $H(t)$ has to be assessed {\it while\/} solving the whole problem: the option price must always be greater or equal than the pay-off function, otherwise you can readily buy the option and execute it just afterwards, thus obtaining risk-free earnings. Therefore $H^{\pm}(t)$ must be settled in such a way that
\begin{equation}
C^{\pm}(S,t)\geqslant P^{\pm}(S,t),
\label{continuity}
\end{equation}
holds~\cite{RCM73}. We must stress again that Eq.~(\ref{option_price}) leads to the right option price from the financial point of view if we use a risk-neutral measure. Under such market measure, and when the underlying stock pays no dividend, we must have
$\EE[Y(S,t) e^{-r(t-t_0)}|\FF(t_0)]=Y(\SI,t_0)$,
for any quoted quantity in the market. In particular
$\EE[C(S,t) e^{-r(t-t_0)}|\FF(t_0)]=C(\SI,t_0)$,
which is true by construction, and
$\EE[S(t) e^{-r(t-t_0)}|\FF(t_0)]=\SI$.
This {\it martingale\/} condition~\cite{CT04} for the risk-neutral evolution of the asset 
implies that our propagator must fulfil that:
\begin{equation*}
\hat{\tilde p}(\omega=-i,s+r)= s^{-1}.
\end{equation*}
If we put this expression into Eq.~(\ref{hat_tilde}) we will arrive to the 
conclusion that waiting times ought to be exponentially distributed, $\psi(t)=\lambda e^{-\lambda t}$, with \begin{equation}
\lambda=\frac{r}{\tilde{h}(\omega=-i)-1}.
\label{lambda}
\end{equation}
Let us analyse the consequences of our previous results. The physical measure and the risk-neutral measure may differ, but they must describe the same kind of process. This means that, if the number of actual market transactions does not follow a Poisson distribution, we will not be able to define a risk-neutral market measure, and the market will become {\bf inefficient}, because then the CTRW has memory and it is not Markovian.
Conversely, if waiting times {\it have\/} an exponential PDF, the risk-neutral market measure can be obtained after imposing Eq.~(\ref{lambda}). Note however that Eq.~(\ref{lambda}) {\it is not\/} a definition for $\lambda$, but a constraint that involves all the physical parameters, as we will show below in a practical example. This is a typical feature of {\bf incomplete} market models.

\section{Perpetual American options and survival probabilities}
\label{Sect_AOSP}

Perpetual 
options, i.e. derivatives with $T \rightarrow \infty$, are not actual traded contracts. However, they 
have practical interest because they represent the limiting value of a far-from-maturity contract, and therefore they may help in the pricing process if the theoretical price cannot be computed~\cite{BAW87} ---which is the most common situation for American options. Moreover, from a pure academic point of view, 
perpetual American options bring a major simplification to the issue: the optimal exercise boundary 
is constant
$H^{\pm}(t)=\HI^{\pm}$,
given that the problem is stationary. Then we have:
\begin{equation*}
C^{\pm}(\SI,t_0) =\EE \left[P^{\pm}(S;\tpm) e^{-r(\tpm-t_0)}|\FF(t_0)\right]. 
\end{equation*}
Observe that, since the process is discontinuous, the actual value of the stock price when we execute the option call (put) will be greater (lower) than $\HI^{\pm}$. 
Therefore $P^{\pm}(S;\tpm)$ is random in general. 
This is not the case for binary option prices, $D^{\pm}(\SI,t_0)$, 
because $P^{\pm}(S;\tpm)=1$ and $\HI^{\pm}=\KI$: 
\begin{equation*}
D^{\pm}(\SI,t_0)=\EE[e^{-r(\tpm-t_0)}|\FF(t_0)]=1-r\widehat\Phi_{\KI} ^ \pm  \left( {s = r;x_0 } \right). 
\end{equation*}
Here we have introduced the complementary 
distribution function of $\tau^\pm$,
\begin{equation*}
\Phi_{K_0}^\pm  \left(t - t_0 ;x_0 \right) = \Pr \left\{ {t_0  < t < \tau^\pm |X(t_0 ) = x_0 =\ln(\SI)} \right\},
\end{equation*}
which, within the CTRW formalism, is usually referred as the survival probability (SP),
$S_{[a,b]} (t - t_0 ;x_0 )$. The SP measures the likelihood that a process beginning in a given region $x_0 \in [a,b]$ remains there all the time interval $t-t_0$.
In all the cases we are dealing with, one of the boundaries of the region is infinite. However, from a practical point of view, it is better to obtain the result for a finite region first, and compute the right limits afterwards. 

The Laplace transform of the SP fulfils by itself 
a renewal equation~\cite{MM07} that
cannot be solved for a general kernel $h(x)$. Therefore, we need to specify a functional form for the jump PDF, and we have focused our attention on the asymmetric two-sided exponential case,
\begin{equation}
h(x)=\frac{\gamma \rho}{\gamma + \rho} \left[e^{-\rho x} \uu_{x\geqslant 0} + e^{\gamma x} \uu_{x<0}\right].
\label{h}
\end{equation}
This model is very suitable for our purposes for two main reasons. On the one hand, we can eventually recover the \BS results in a certain limit. And, on the other hand, 
the solution to this problem is intricate but obtainable,
\begin{eqnarray*}
\hat{S}_{[a,b]}(s;x_0)&=&\frac{1}{s}\left\{1 -\frac{\hat{\psi}(s)}{\Delta} e^{\betap x_0} \left[\rho(\gamma + \betam)e^{-\betam a} - \gamma (\rho -\betam)e^{-\betam b}\right] \right. \nonumber \\ 
&-& \left. \frac{\hat{\psi}(s)}{\Delta} e^{\betam (x_0-a-b)} \left[\gamma (\rho -\betap)e^{\betap a} - \rho(\gamma + \betap)e^{\betap b}  \right] \right\},
\end{eqnarray*}
where we have defined some auxiliary magnitudes:
\begin{eqnarray*}
\betap&=&\frac{\rho-\gamma}{2}+\sqrt{\left(\frac{\rho+\gamma}{2}\right)^2-\gamma \rho \hat{\psi}(s)}, 
\quad \betam=\frac{\rho-\gamma}{2}-\sqrt{\left(\frac{\rho+\gamma}{2}\right)^2-\gamma \rho \hat{\psi}(s)}, \\
\Delta&=&(\rho-\betap)(\gamma + \betam) e^{(\betap -\betam)a} - (\rho-\betam)(\gamma + \betap) e^{(\betap -\betam)b}.
\end{eqnarray*}
We must remember, however, that we only need the value of $\hat{S}_{[a,b]}(s;x_0)$ when either $a \rightarrow -\infty$ or $b \rightarrow +\infty$, what simplifies the final expressions:
\begin{eqnarray*}
\hat \Phi_{\KI}^+ (s;x_0 ) = \frac{1}{s} -\frac{\gamma}{s}\frac{{\hat \psi (s)}}{\gamma +\betap} \left(\frac{\SI}{\KI}\right)^{\betap}, \quad
\hat \Phi_{\KI}^- (s;x_0 ) = \frac{1}{s} - \frac{\rho}{s}\frac{{\hat \psi (s)}}{\rho-\betam} \left(\frac{\SI}{\KI}\right)^{\betam}.
\end{eqnarray*}
We can compare these results with the same outcomes for the Wiener process:
\begin{eqnarray*}
\hat \Phi_{\KI}^+ (s;x_0 ) = \frac{1}{s}\left\{1 - \left(\frac{\SI}{\KI}\right)^{\overline{\betap}} \right\}, \quad 
\hat \Phi_{\KI}^- (s;x_0 )  = \frac{1}{s}\left\{1 - \left(\frac{\SI}{\KI}\right)^{\overline{\betam}} \right\}, 
\end{eqnarray*}
with
\begin{equation*}
\overline{\betap}=-\frac{\vartheta}{\sigma^2}+\frac{1}{\sigma^2}\sqrt{\vartheta^2 + 2\sigma^2 s}, \quad 
\overline{\betam}=-\frac{\vartheta}{\sigma^2}-\frac{1}{\sigma^2}\sqrt{\vartheta^2 + 2\sigma^2 s}. 
\end{equation*}
Here $\sigma$ is the volatility, the square root of the diffusion coefficient, and $\vartheta$ is the drift of the diffusive process.
The two formulas coincide for small values of the mean sojourn time $\mu$, for which we can expand $\hat \psi (s)\sim 1-\mu s$, and for large values of $\rho$ and $\gamma$, 
once one identifies $\vartheta=(\gamma  - \rho )/\gamma \rho \mu $, and $\sigma ^2=2/\gamma \rho \mu$. 

\section{Risk-neutral prices} \label{Sect_Risk_Neutral}
In this section we will present results for the jump PDF in Eq.~(\ref{h}) when we have adopted a risk-neutral market measure, that is $\psi(t)=\lambda e^{-\lambda t}$. We will proceed by replacing $\lambda$ according to Eq.~(\ref{lambda}). 
This choice is feasible only when $\gamma>\rho-1>0$, because we need $\tilde{h}(\omega=-i)$ to be bounded,   
and $\lambda$ must be positive definite:
\begin{equation*}
\lambda=r\frac{(\rho-1)(\gamma+1)}{\gamma-\rho+1}.
\end{equation*}
The above conditions simplify the values of $\betap=1$ and $\betam=-(\gamma-\rho+1)$, after we have set $s=r$, and lead to the following expressions for the {\it live\/} binary call ($\SI\leqslant \KI$) and put ($\SI\geqslant \KI$) option prices:\footnote{We will drop hereafter the dependence on $t_0$, because it is merely formal.}
\begin{equation*}
D^{+}(\SI)=
\frac{\rho-1}{\rho}\frac{S_0}{K_0}, \qquad D^{-}(\SI)=
\frac{\rho-1}{\gamma}\left(\frac{K_0}{S_0}\right)^{\gamma-\rho+1}.
\end{equation*}
Note that in both cases the prices are discontinuous, 
since $D^{+}(\SI>\KI)=D^{-}(\SI<\KI)=1$. This feature 
disappears when considering continuous trading, $\lambda\rightarrow \infty$. We can approach to this limit by letting $\rho \rightarrow \infty$ and $\gamma \rightarrow \infty$, but in such a way that the difference remains finite $\infty>\gamma - \rho+1=\varepsilon>0$. In fact, in this case we recover the \BS prices~\cite{McK65,RR91}, since then $\betap  = \overline \betap   = 1$, and $\betam  = \overline \betam   =  - \varepsilon  =  - 2r/\sigma^2$.

\begin{figure}[hbtp]
\begin{center}
\includegraphics[width=0.80\textwidth,keepaspectratio=true]{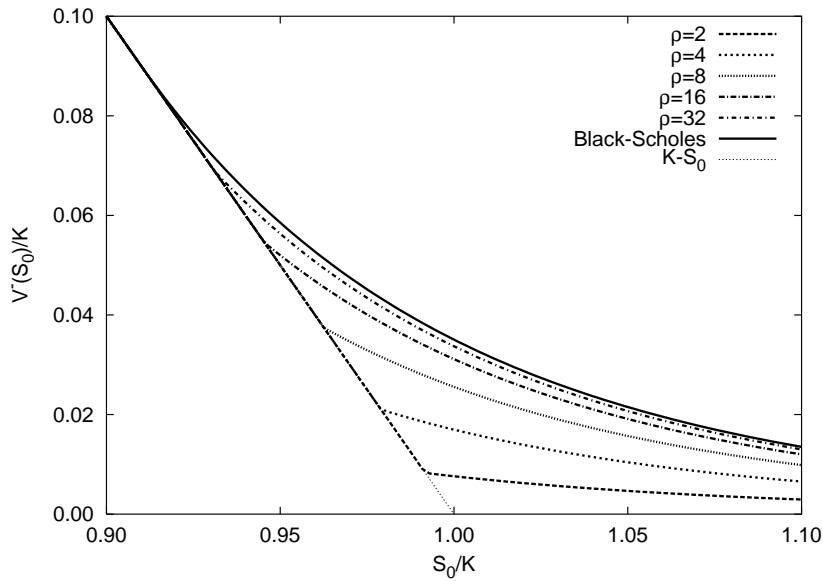} \caption{Option prices for different values of $\rho$. We show several plots of {\it vanilla\/} puts in terms of the moneyness $S_0/K$, after setting $\gamma=\rho-1+\varepsilon$, and $\varepsilon=10$. We have also represented the \BS price for $r=5\%$ and $\sigma=10\%$.} \label{VanillaPlot}
\end{center}
\end{figure}

We can now consider the usual {\it vanilla\/} pay-off which we introduced in Eq.~(\ref{payv}). The problem here is that, as we advanced in Section~\ref{Sect_AOSP}, $P^{\pm}(S;\tpm)$ is a random magnitude. 
When $h(x)$ follows Eq.~(\ref{h}), however, $\EE[P^{\pm}(S;\tpm)|\tpm]$
does not depend on $\tpm$. 
Then the {\it vanilla\/} option price $V^{\pm}(\SI)$ can be easily computed:
\begin{eqnarray*}
V^{+}(\SI)=S_0-\frac{\rho-1}{\rho}\frac{\SI}{\HI^+}K&\quad&(\SI  \leqslant \HI^+ ),\\ 
V^{-}(\SI)=\left(K-\frac{\gamma}{\gamma+1}\HI^-\right)\frac{\rho-1}{\gamma}\left(\frac{\HI^-}{S_0}\right)^{\gamma-\rho+1}&\quad&(\SI  \geqslant \HI^- ),
\end{eqnarray*}
in terms of $\HI^{\pm}$. When we impose condition~(\ref{continuity}) on $V^{+}(\HI^+)$ we obtain an inconsistent expression for finite values of $\HI^+$ and the option is never exercised, $V^{+}(\SI)=S_0$. When we demand $V^{-}(\HI^-)=(K-\HI^-)$, in turn, we get
\begin{equation*}
\HI^-=\frac{(\gamma+1)(\gamma-\rho+1)}{\gamma(\gamma-\rho+2)}K <K.
\end{equation*}
The continuous-trading limit leads again to the \BS formula~\cite{McK65,IJK90}, 
as it can be observed in Fig.~\ref{VanillaPlot}. There we represent put option prices for different values of $\rho$ and $\gamma$. In order to reach the diffusive limit for a given choice of $r$ and $\sigma$, we have imposed the relationship $\gamma=\rho-1+2r/\sigma^2$.

\section{Conclusions} \label{Sect_Conclusions}

We have argued for the convenience of the use of CTRW's in the modelling of stochastic processes in finance. CTRW is a well suited tool for representing market changes at the lower time scale, the transaction-to-transaction evolution. We have shown how CTRW-based results can be easily adapted to financial terminology with little effort. In particular, we have analysed their possible applications in option pricing problems, by relating the computation of survival probabilities with the price of perpetual American derivatives for the two most ubiquitous flavours: binary options and {\it vanilla\/} options. 

We have considered risk-neutral scenarios for which we have obtained results that are fair from the financial point of view. Moreover, we have shown how classical \BS results can be obtained from these expressions under certain limits. Finally, we have pointed out that our approach may be fruitful in the future in exploring a particular class of inefficient market models. 
There are multiple evidences~\cite{ES06,MMP05} pointing to the fact that transactions are not exponentially distributed. 
In such a case the CTRW machinery may be very helpful in providing solutions for a general pausing time distribution. 

\ack We acknowledge support from MEC under contract No. FIS2006-05204-E.

\end{document}